\documentclass{iopconfser}
\usepackage{graphics}
\usepackage{color}
\usepackage{hyperref}
\usepackage{tikz}
\usepackage{tikz-feynman}
\usepackage[bitstream-charter]{mathdesign}
\begin{document}

\title{\bf Irreducible Multi-Particle Representations of the Poincar\'e Group as a Basis for the Standard Model}

\author{Walter Smilga}

\affil{Munich, Germany}

\email{wsmilga@compuserve.com}

\begin{abstract}
A phenomenological description of the Stern--Gerlach experiment yields a mathematical structure equivalent to that of a spin-½ particle, described by an irreducible unitary representation of the Poincar\'e group.
In the corresponding irreducible two-particle representation, two-particle states have the form of an integral over product states. They describe a correlation between the particles with the structure of the electromagnetic interaction and a coupling constant that numerically equals the electromagnetic coupling constant.
This coupling constant is essentially the normalisation factor of these two-particle states.
The Standard Model of particle physics describes the electromagnetic interaction by a perturbation algorithm, where the experimental value of the electromagnetic coupling constant is inserted by hand.
It is argued that it does not make sense to insert a normalisation factor without checking the range of integration of the corresponding integral and adjusting it if necessary.
This adjustment provides the perturbation algorithm with the mathematically consistent structure of a non-local, relativistic, two-particle quantum mechanics.
Similarly, multi-particle representations determine a gravitational interaction that, in the quasi-classical limit, is described by the field equations of conformal gravity. 
A calculated, galaxy-specific value of the gravitational constant matches the experimental value.
\end{abstract}
 
\section{Introduction}
\label{sec:intro}

The Standard Model of particle physics, as we know it today, describes three of the four fundamental interactions: the electromagnetic, the weak, and the strong interaction. 
So far, it has failed to describe the gravitational interaction. 
It also does not provide the numerical values of the coupling constants and masses. 
These parameters still need to be entered by hand as empirical values.

The development of the Standard Model has stagnated for several decades, which is not least due to the fact that the higher order terms of the perturbation expansion exhibit inconsistencies, in the form of divergent integrals.
In practice, these integrals are regularised by using a renormalisation procedure. 
This procedure conceals the inconsistencies, but does not eliminate them.

A fundamental weakness of the Standard Model is that it is not based on quantum-specific experiments, but rather on the firmly established opinion that the interaction between elementary particles must be described by locally interacting quantum
fields.\footnote{Werner Heisenberg in his 1961 lecture on particle physics: ``Today we know that a theory of elementary particles must be a local quantum field theory.''}
There are good reasons to question this opinion, given that quantum theory is essentially a non-local theory. 
This fact was addressed in 1935 by the Einstein--Podolsky--Rosen paradox \cite{epr}.
We have since learned that this particular type of non-locality is fully compatible with relativistic causality.
In fact, there is widespread agreement that the requirement of locality is directly responsible for the divergences mentioned.
Therefore, a critical review of the role of locality within the Standard Model is overdue.

A suitable basis for such a review is the Stern--Gerlach experiment \cite{sg}, which has not yet received the attention it deserves.
Although Stern and Gerlach carried out this experiment in 1922, it was not until 1925 that Uhlenbeck and Goudsmit correctly attributed it to the spin of the electron.
Since then, the Stern--Gerlach experiment has been regarded as the experiment that proved the directional quantisation of spin.
In the following, it will play a central role in `debugging' the Standard Model.

\section{Stern--Gerlach experiment}
\label{sec1}

In the Stern--Gerlach experiment, we observe the spin of the outer electron of a silver atom in an inhomogeneous magnetic field.
We find that the spin has only two orientations relative to the measuring device.
If we repeat the measurement with a second Stern--Gerlach device and rotate this device by an angle $\beta$ before the second measurement is done, the result will be non-deterministic; but if we repeat the measurement with different angles, we will observe a pattern. 
This pattern describes a probability distribution that depends only on the result of the first measurement and the angle of rotation of the second device. 

What we see is a typical quantum phenomenon. 
But this is not due to any quantisation rules: it is just because spin has only a discrete degree of freedom. 
Instead of trying to “explain” this experimental fact, it is more beneficial to study its consequences.

\section{Mathematics of the Stern--Gerlach experiment}
\label{sec2}

To do this, we need a mathematical description of the observed pattern \cite{ws1}.
Obviously, one can choose a two-dimensional complex vector space to encode the information about the initial measurement as a two-dimensional vector $\Psi$. 
To encode the information about the rotation by the angle $\beta$, we rotate the vector $\Psi$ to $\Psi'$. 

The probability $p$ of measuring the original orientation (or the opposite) in the second measurement is then given by Born’s rule
\begin{equation}
p=\left| \left<\Psi'\mbox{\textbar}\Psi\right> \right|^2 .
\end{equation} 
This rule either follows from Gleason’s theorem \cite{gt} or directly from the experimental data.

The Stern--Gerlach experiment thus provides the experimental basis for the quantum mechanical notion of `spin'. 
We can say: spin describes what a classical observer sees when observing a bit-valued entity. 
In more mathematical terms: spin is the covariant description of a bit-valued entity relative to an observer that is subject to rotations.
This description is based on a representation of the rotation group on a two-dimensional complex vector space.
Note that, in this context, the term `observer' refers to a classically described experimental or naturally given set-up that interacts with the quantum object in some way.

\section{The covariant concept of a particle}
\label{sec3}

Since observers can move freely through space and time, we must extend the description of spin by replacing the representation of the rotation group with that of a representation of the Poincaré group \cite{ew} in order to fully comply with Poincar\'e symmetry.
Such representations are provided by the solutions of the Dirac equation \cite{pamd}
\begin{equation}
\left(i\gamma^\mu \frac{\partial}{\partial x^\mu} - m \right) \psi(x) = 0.              						 \label{2-4}
\end{equation}
The matrices $\gamma^\mu$ are $4\times4$ Dirac matrices.
The Dirac equation admits plane wave solutions of the form
\begin{equation}
\psi(x) = e^{-i p \cdot x/\hbar}\, u(p),																\label{2-5}
\end{equation} 
where $u(p)$ is a four-component spinor which satisfies 
\begin{equation}
\left(\gamma^\mu p_\mu - m\right) u(p) = 0.														\label{2-6}
\end{equation}
This gives the spin a four-momentum $p$, which satisfies the mass shell relation 
\begin{equation}
p^\mu p_\mu = m^2. 																			\label{2-8}
\end{equation}

We have thus created an operational description of spin-½ particles that establishes a transparent connection between experiment and theory.
This description takes into account only the experimental facts of the Stern--Gerlach experiment and Poincaré invariance. 
It describes a spin-½ particle in terms of information about preparing and observing a bit-valued entity in spacetime.\footnote{This can be seen as the realisation of Wheeler's idea of ``it from bit'' \cite{jaw}}.

Note that this description contains no indication of an electrical charge. 
Instead, the properties of the particle relative to its environment are encoded using the state of an irreducible representation of the Poincaré group.
When talking about the behaviour of this particle in spacetime, we are actually talking about the evolution of information in spacetime.
 
This concept of a particle can easily be extended to two- and multi-particle configurations by forming tensor products of the individual state spaces. 
Similar to single-particle representations of the Poincar\'e group, the resulting representations can be classified by the eigenvalues of two Casimir operators, corresponding to the mass and angular momentum of the configuration. 
This classification results in a spectrum of irreducible representations that differ in mass and angular momentum.
Different masses mean that these representations exhibit non-trivial energetic properties. 
These properties can manifest as interactions, as will be shown in the following.

\section{Two-particle configurations}
\label{sec4}

At GROUP32 in 2018, I reported on the structure of two-particle representations \cite{ws2}. See also \cite{ws3}.

From product states, we can form superpositions, for example, eigenstates of the total linear and angular momentum 
\begin{equation}
\left.\mbox{\textbar}\mathbf{p},l\right> = \int_\Omega \! \mathrm{d}s \;
c_\mu({\mathbf{p}},l,{\mathbf{k}(s)})  \,
\left|\textstyle{\frac{1}{2}}\mathbf{p}+\mathbf{k}(s)\right>\gamma^\mu\left|\textstyle{\frac{1}{2}}\mathbf{p}-\mathbf{k}(s)\right> .  \label{4-1}
\end{equation}
The angular momentum consists of spin and orbital angular momentum, the latter of which is defined by the motion of the particles relative to each other.
The integral is over a circular domain $\Omega$ on the two-particle mass shell, defined by the mass-shell relations of the four-momenta $p_1$, $p_2$ and $p = p_1 + p_2$,
\begin{equation}\vspace{0.01cm}
 p^2 = M^2, \; \;  p_1^2 = p_2^2 = m^2. 									\label{4-2}
\end{equation}
The state (\ref{4-1}) must be normalised by the inverse of the square root of the volume of Omega, 
\begin{equation}
\omega = V(\Omega)^{-1/2} . 												\label{4-3}
\end{equation}
If we extract the normalisation factor $\omega$ from the coefficients $c_\mu$, these coefficients become essentially phase factors with $c^*c =1$.
The normalised state is then written as $\omega\!\left.\mbox{\textbar}\mathbf{p},l\right>$.

By relation (\ref{4-3}), the numerical value of the normalisation factor $\omega$ is inseparably linked to the domain of integration $\Omega$.
In the following, this numerical value will help to determine the correct domain of integration when this is called into question.

In \cite{ws3}, a gedanken experiment was described: a scattering experiment with two incoming particles with momenta $\mathbf{p_1}$ and $\mathbf{p_2}$, the two-particle eigenstate from the above as the intermediate state, and two outgoing particles with momenta $\mathbf{q_1}$ and $\mathbf{q_2}$. 
The scattering amplitude 
\begin{equation}
S = 4 \pi \omega^2\left<\mathbf{p}_1,\mathbf{p}_2\, \mbox{\textbar}\, \mathbf{p},l \right>\!\!
\left<\mathbf{p},l \,\mbox{\textbar}\, \mathbf{q}_1,\mathbf{q}_2\right>   \nonumber
\end{equation}
is then essentially determined by the constant $4\pi\omega^2$. 
The numerical value of this constant has been calculated with the result $1/137.036$.\footnote{This calculation verifies the semi-empirical Wyler formula, as originally proposed by Wyler \cite{aw}.} 
This value coincides with the experimental value (CODATA $1/137.035999$) of the electromagnetic fine-structure constant $\alpha$. 

This implies that forming a two-particle eigenstate of total momentum and angular momentum results in an interaction between the particles that exhibits the characteristics of an electromagnetic interaction.
However, this interaction is not a property of the individual particles, which, by definition, do not carry a charge. 
Rather, it emerges when such two-particle states are formed.

The action of the Poincar\'e group on such an eigenstate generates the state space of an irreducible unitary two-particle representation of the Poincar\'e group.
The numerical value of the fine-structure constant can therefore be considered as a signature of such a representation.

\section{Comparison with the Standard Model}
\label{sec5}

The Standard Model describes a similar amplitude, the amplitude of electron--electron scattering in the second order of the perturbation series.
This amplitude corresponds to this Feynman diagram\\
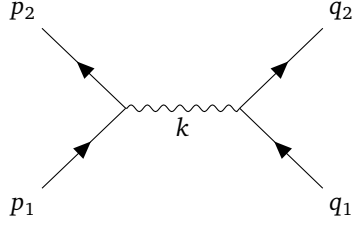
 \begin{center}
\begin{tikzpicture}
  \begin{feynman}
    \vertex (b);
    \vertex [right=of b] (c);
    \vertex [below left=of b] (d) {\(p_1\)};
    \vertex [above left=of b] (a) {\(p_2\)};
    \vertex [above right=of c] (e) {\(q_2\)};
    \vertex [below right=of c] (f) {\(q_1\)};

    \diagram* {
      (d) -- [fermion] (b) -- [fermion] (a),
      (f) -- [fermion] (c) -- [fermion] (e),
      (b) -- [boson, edge label'=\(k\)] (c), 
    };
  \end{feynman}
\end{tikzpicture}  
\end{center}
Application of the Feynman rules in momentum form (see \cite{sss}, pp.\ 478--479) leads to the amplitude
\begin{equation}
R = -\frac{i}{\pi}\frac{e^2}{4 \pi} \int\!d^4 k\,d^4 k' \dots \nonumber  
 \tilde{w}(\mathbf{p}_2)\, \gamma^\mu w(\mathbf{p}_1) \; \tilde{w}(\mathbf{q}_2)\, \gamma_\mu w(\mathbf{q}_1) 
\dots  \frac{1}{ k^2+i \epsilon} . 							\label{5-1}
 \vspace{-0.05cm}
\end{equation}
The dots stand for delta-functions that ensure the conservation of momentum. 
 
We see two vertices, $\tilde{w}(\mathbf{p}_2)\, \gamma^\mu w(\mathbf{p}_1)$ and $\tilde{w}(\mathbf{q}_2)\, \gamma_\mu w(\mathbf{q}_1)$, with the same structure as in the two-particle state discussed above. 
We also find the corresponding integrals and the normalisation factors, contained in the term $e^2 / 4\pi$.
Note that the numerical value of this term is not determined by the Standard Model. 
Instead, here the value of the squared empirical coupling constant, i.e.\,the fine-structure constant $\alpha$, is inserted by hand.
Numerically, this value equals $4\pi \omega^2$, where $\omega$ is the normalisation factor of a two-particle state of an irreducible two-particle representation of the Poincar\'e group. 
It is evident that this value does not normalise the integrals over $k$ and $k'$ over the entire four-dimensional momentum space $\mathbb{R}^4$.
This clearly indicates an inconsistency in the configuration of the perturbation algorithm.

The domain of integration $\mathbb{R}^4$ results from the local structure of the interaction term of quantum electrodynamics, $e\,  \psi(x) \gamma_\mu \psi(x)\,  A^\mu(x)$, which in the momentum representation corresponds to a convolution integral with the domain of integration $\mathbb{R}^4$.
In the scattering amplitude (\ref{5-1}), the $\delta$-functions $\delta^{(4)}(p_2-p_1+k))$ and $\delta^{(4)}(q_2-q_1-k')$ restrict $k$ and $k'$ to the differences of the incoming and outgoing momenta, so that here $\mathbb{R}^4$ does not become effective.

The situation is different for Feynman diagrams containing a loop (see, for example, \cite{sss}, pp. 508--524), where the internal momenta are not linked to either the incoming or outgoing momenta.
This is where $\mathbb{R}^4$ comes into play, resulting in divergent integrals.

Given the relation (\ref{4-3}), it is meaningless to enter the experimental value of the fine-structure constant manually without checking and adjusting the corresponding domain of integration if necessary.

Replacing the domain of integration $\mathbb{R}^4$ with $\Omega$ gives the perturbation algorithm a mathematically consistent structure based on relativistic (non-local) quantum mechanics. 
More precisely, it is based on an irreducible unitary two-particle representations of the Poincaré group.  

Inspection of the divergent integrals of the perturbation series reveals that this replacement `regularises' these integrals in a manner similar to a cut-off.
However, unlike a cutoff, it does not destroy Poincar\'e invariance.

\section{Configurations of $N$ spinless particles}
\label{sec6}

Two years ago, at QTS12, I reported on the structural properties of $N$-particle representations of the Poincaré group \cite{ws4}, with $N$ equal to $2.4\times10^{67}$; this is the estimated number of atoms in our 
galaxy\footnote{Steve Cavill on {\it Oxford Education Blog}: ``Our galaxy, the Milky Way, contains approximately 100 to 400 billion stars. If we take this as 200 billion or $2 \times 10^{11}$ stars and assume that our sun is a reasonable average size we can calculate that our galaxy contains about $(1.2 \times 10^{56}) \times (2 \times 10^{11}) = 2.4 \times 10^{67}$ atoms.''}. 
A state of this representation is characterised by fixed total linear and orbital angular momenta. 
As an eigenstate of the orbital angular momentum this state has a rotational symmetry. 

In the quasi-classical limit of large quantum numbers, the rotational symmetry of this state is preserved. 
Therefore, the total momentum $p$ is a sum over time-dependent particle momenta
\begin{equation}
\sum_{i=1}^{N}\; p_i(t) = p .
\end{equation}
The structure of this equation is form-invariant under particle-individual Lorentz transformations and also under conformal scaling of the momenta, that is, individual scaling of the particle masses.

A consequence of the conservation of the total momentum is that the curvature of the trajectory of a test particle is determined by the time-dependent momenta of the other particles. 
So, there must be a general relation between the curvature of this trajectory and the energy-momentum tensor of the $N$-particle configuration. This strongly reminds us of Einstein's theory of gravity \cite{ae}.

In formulating this theory, Einstein required that the general laws of nature be given a form that is covariant under the relevant symmetry operations; here, these symmetries include conformal scaling. 

The energy-momentum tensor $T^{\mu\nu}$ is not covariant under conformal scaling.
However, the traceless energy-momentum tensor, $T^{\mu\nu} - \textstyle{\frac{1}{4}} g^{\mu\nu}\,T$,  where $T \equiv T^\mu_\mu$, is covariant. 
A corresponding traceless curvature tensor $W^{\mu\nu}$ can be constructed from the conformal-invariant Weyl tensor $C$, which is the traceless part of the Riemann curvature tensor. 
A variational principle then leads to the field equations of conformal gravity \cite{pm}
\begin{equation}
W^{\mu\nu} = \kappa\, (T^{\mu\nu} - \textstyle{\frac{1}{4}} g^{\mu\nu}\,T)	
\end{equation}
with
\begin{equation}
W^{\mu\nu} = 2 \,C^{\mu\lambda\nu\xi}_{\;\;\;\;\;\;\;\;\;\;\;;\lambda;\xi} - C^{\mu\lambda\nu\xi} R_{\lambda\xi}. 
\end{equation}
Here, $R_{\lambda\xi}$ is the Ricci tensor. 
The constant $\kappa$ is needed for dimensional reasons. 
It has the same dimension as $1/(h c)$, where $h$ is  Planck's constant and $c$ the speed of light.

A simple consideration leads to an estimate of the strength of the gravitational interaction between two given particles.
If the momentum of particle $1$ changes by a momentum $\Delta p$, then the probability that this change is compensated for by a change of the momentum of particle 2 is $1/N$, where $N$ is the total number of particles of the configuration.
A reasonable estimate of the numerical value of $\kappa$ is therefore given by 
\begin{equation}
1 / (N\,h\,c) \approx 2.1 \times10^{-43}\;\mathrm{kg^{-1} m^{-3}\,s^2},
\end{equation}
where $N$ is the estimated number of atoms in our galaxy.
This value is almost identical to the corresponding coupling constant 
\begin{equation}
8\pi G/c^4 \approx 2.077 \times 10^{-43}\;\mathrm{kg^{-1} m^{-1}\,s^2}      
\end{equation}
in Einstein’s field equations. 
Note that the dimensions of these constants are slightly different. 

To understand this difference, consider the field equations of conformal gravity for a rotationally symmetric system.
The gravitational field $\Phi(r)$ can then be decomposed into terms of order $1/r$ and $r$ (see \cite{pm2})
with dimensioned coefficients $c_1$ and $c_2$,
\begin{equation}
\Phi(r) = c_1 O(1/r) + c_2 O(r) .
\end{equation}
For small values of $r$, compared to the radius of the galaxy, the decomposition is dominated by $O(1/r)$, meaning that $c_1$ takes on a value of $1$.
As we know from Einstein's field equations, this term is described by the Ricci tensor.
For small $r$, the field equations therefore take the form of Einstein's equations,
\begin{equation}
c_1\, R^{\mu\nu} \approx \kappa\; T^{\mu\nu} \;\; \mbox{ or } \;\; R^{\mu\nu} \approx \frac{\kappa}{c_1} T^{\mu\nu}.
\end{equation}  
The coupling constant $\kappa/c_1$ has the same numerical value and the same dimension as the coupling constant in Einstein’s equations. 
This allows us to compare these values and conclude that they are in excellent agreement.

\section{Conclusions}
\label{sec7}

The analysis presented above reveals the fundamental relevance of the Stern–Gerlach experiment to relativistic quantum mechanics. As a representation of a basic quantum mechanical measurement process, this experiment shows how nature resolves the conflict between discrete and continuous degrees of freedom, thus providing a comprehensive understanding of quantum mechanics at its most fundamental level. Furthermore, this experiment provides the basis for an operational concept for spin-½ particles, in which irreducible unitary representations of the Poincar\'e group are used to encode information about the preparation of these particles.

Irreducible two-particle representations are then used to determine the electromagnetic interaction together with the value of the electromagnetic coupling constant, $\alpha$. This result is empirically supported by the fact that the experimentally determined value of $\alpha$ matches the calculated value.

The recognition that $\alpha$ is essentially the normalisation factor for two-particle states makes a significant contribution in terms of understanding and correcting the inconsistencies in the perturbation algorithm. These inconsistencies are caused directly by the local structure of the interaction term. As evidenced by the divergent integrals in the perturbation algorithm, this locality is not compatible with the structure of two-particle states, a structure that is determined by Poincar\'e symmetry.
Correcting these integrals, as indicated by the value of the fine-structure constant, leads to a new understanding of the Standard Model within a consistent framework of relativistic, non-local, particle-based quantum mechanics.

An irreducible representation describing $N$ spinless particles determines a quantum gravity that, in the quasi-classical limit, is described by the field equations of conformal gravity with a galaxy-specific gravitational constant, ~$\kappa$. These field equations determine a Riemannian manifold that can be identified with the empirical spacetime manifold. Its curvature reflects the rotational symmetry of the eigenstates of the orbital angular momentum. Again, this result is empirically supported by the fact that the experimental value of $\kappa$ matches the calculated value.

The value and dimension of the gravitational constant indicate that Einstein's theory of gravity is an approximation to conformal gravity, and is valid for local gravitational phenomena within a galaxy.

Conformal gravity correctly describes the observed structures of galaxies without the need for dark matter \cite{pm2}. The galaxy-dependence of the gravitational constant is expected to have significant implications for cosmological models describing the evolution of galaxies.

\end{document}